\documentstyle[epsfig,12pt]{article}

\newcommand{\nc}{\newcommand}
\nc{\beq}{\begin{equation}} \nc{\eeq}{\end{equation}}
\nc{\beqa}{\begin{eqnarray}} \nc{\eeqa}{\end{eqnarray}}

\begin{document}

\title{{\large {\bf Holography, Entropy and Extra Dimensions}}}
\author{Deog Ki Hong\thanks{dkhong@pnu.edu} \\
Department of Physics\\
Pusan National University, Pusan 609-735, Korea \\
\\
Stephen D.H.~Hsu\thanks{
hsu@duende.uoregon.edu} \\
Department of Physics \\
University of Oregon, Eugene OR 97403-5203 \\}
\maketitle

\begin{abstract}
We show that higher dimensional models (brane worlds) in which the
scale of quantum gravity $M_*$ is much smaller than the apparent
scale $M_P \sim 10^{19}$ GeV violate the covariant entropy bound
arising from holography. The thermodynamic entropies of
astrophysical black holes and sub-horizon volumes during big bang
nucleosynthesis exceed the relevant bounds unless $M_* >
10^{(4-6)}$ TeV, so a hierarchy relative to the weak scale is
unavoidable. We discuss the implications for extra dimensions as
well as holography.
\end{abstract}

\newpage

Recent progress in string theory and the study of quantum black
holes strongly suggests that the maximum information content of a
spacetime region is related to its surface
area~\cite{'tHooft:gx,Susskind:1994vu,Bousso:2002ju}.
The idea has its origins in the proposal of Bekenstein that the
area of a black hole is proportional to its entropy \cite{Bekenstein:ur},
and that black holes obey a generalized second law of
thermodynamics (GSL) \cite{Bekenstein:ax}. A covariant generalization of
these ideas \cite{Fischler:1998st,Bousso:1999xy} has passed a number of theoretical
tests, and implies a deep relationship between geometry and
information which arises due to quantum gravity.

At first glance, the proposal that the maximum information in a volume only
grows as the surface area seems obviously false. At least, it contradicts
the usual counting used in statistical physics or quantum field theory.
However, 't~Hooft~\cite{'tHooft:gx}
showed that if configurations in the volume which would
have already led to gravitational collapse are excluded, the number of
states grows less rapidly than the exponential of the area, or equivalently
that the entropy is bounded above by the area, in Planck units. We give a
version of 't~Hooft's argument below, generalized to D spacetime dimensions.

Models with $D > 4$ dimensions, in which ordinary matter is
constrained to a 3+1 dimensional subspace (the brane), while
gravity propagates in all D dimensions, allow for the possibility
that the fundamental scale of gravity, $M_*$, is much smaller than
the {\it apparent} Planck scale, $M_P = 10^{19}$ GeV, governing
gravitational interactions on the
3-brane~\cite{Arkani-Hamed:1998rs,Randall:1999ee}. If $M_*\sim$
TeV, these models solve the hierarchy problem in a novel manner.
However, the surface entropy or information density is
consequently much smaller than in ordinary theories of gravity. We
refer to models in which the extra dimensional space is flat as
ADD models \cite{Arkani-Hamed:1998rs}, and those with warped extra
dimensions as RS models \cite{Randall:1999ee}.

In this letter we examine systems, such as supernova cores and the
early universe, which are accurately described by ordinary
thermodynamics. In order that these systems not saturate
holographic bounds on their entropy, we deduce a lower bound on
the fundamental scale of quantum gravity: $M_* > 10^{(4-6)} ~{\rm
TeV}$. Clearly, this bound is problematic for brane world
solutions to the hierarchy problem.

\bigskip

\noindent {\bf Notation and Preliminaries}

$M_*$ is the dynamical scale of quantum gravity. In ADD/RS universes it can
be as low as a few TeV. $M_P$ is the apparent Planck scale, or $10^{19}$ GeV.
$D$ is the number of spacetime dimensions. In $D$ dimensions the surface
area (or surface volume, for $D > 4$) of a black hole scales like $R^{D-2}$.

The Einstein action is \beq S ~=~ M_*^{D-2}~\int d^D x~\sqrt{-g}~
{\cal R}~~~ \eeq and the (dimensionless) gravitational potential
per unit test mass behaves as \beq \Phi ~\sim~ {E \over M_*^{D-2}
R^{D-3}} ~\sim~  { ER \over (M_* R)^{D-2}} \eeq Thus, the
Schwarzschild radius in $D$ dimensions is: \beq \label{SR} R_s
\sim ( M_*^{2-D} E )^{1/(D-3)} \eeq

\bigskip
\noindent {\bf Holography in D dimensions}

To generalize 't~Hooft's result~\cite{'tHooft:gx}, we compute the entropy
of a region of size $R$ under the condition that it is on the
verge of collapsing to a black hole, or $R \sim R_s$. We assume a roughly
spherical geometry throughout. As we will see later in the paper,
assuming a brane world geometry modifies the results substantially.

The dominant configurations are thermal\footnote{Temperature here
is fictitious. It lets us characterize the dominant configurations
in phase space, since in thermal equilibrium the entropy
density is maximal for a given energy density.}, characterized by a
temperature $T$, energy density $T^D$ and entropy density
$T^{D-1}$. The total energy and entropy of the region are \beq
\label{ES} E \sim R^{D-1} T^D~,~~~ S \sim R^{D-1} T^{D-1}~~~. \eeq

Substituting (\ref{SR}) into (\ref{ES}), we obtain the bound: \beq
M_*^{2-D} R^{D-1} T^D  ~<~ R^{D-3}~~, \eeq which implies \beq T
~<~ \left( M_*^{D-2} R^{-2} \right)^{1/D}~~, \eeq and the entropy
bound \beq S \sim R^{D-1} T^{D-1} ~<~
M_*^{(D-2)(D-1)/D}~R^{D-1}~R^{-2(D-1)/D}~~, \eeq or \beq S ~<~
R^{D-3+2/D} ~\sim~ R^{(D-2) + (2/D - 1)}~~~ \eeq in appropriate
units. This is always at least as strong as the holographic bound,
which depends on the surface area: $S < R^{D-2}$. The two bounds
coincide when $D=2$. Note that in $D=2$ the boundary of the black
hole is simply two points, so our bound should scale as $R^0$,
which it does. For $D=4$ we obtain 't~Hooft's
result~\cite{'tHooft:gx} that $S < A^{3/4}$. When we allow for an
admixture of black holes in the region, the holographic bound is
recovered. (It is saturated when the region becomes a single black
hole.)

Bousso \cite{Bousso:1999xy,Bousso:2002ju} formulated a covariant
generalization to previous holographic entropy bounds which
addresses some of the failings of the original spacelike bounds
and which applies even in cosmological or strongly gravitating
settings. In what follows we will apply Bousso's covariant entropy
bound (CB) to higher dimensional theories. The bound can be stated
as follows.

{\it Let $A(B)$ be the area of an arbitrary $D-2$ dimensional
spatial surface, which need not be closed. A $D-1$ dimensional
hypersurface $L$ is the light-sheet of $B$ if $L$ is generated by
light rays extending orthogonally from $B$, which have
non-positive expansion everywhere on $L$. Let $S$ be the entropy
on any light sheet of $B$. Then $S \leq {1 \over 4} A(B)$.}

The entropy of a light sheet is given by that of the matter
intersecting the sheet. For simple cases, such as a suitable
closed spacelike surface surrounding a weakly gravitating system,
the covariant bound reduces to the usual area bound.

\bigskip
\noindent {\bf Application to extra dimensions}

Assume an ADD/RS world in which the standard model degrees of
freedom are confined to a 3-brane while the gravitational degrees
of freedom propagate in $D > 4$ dimensions. The large effective
volume $V_w$ of the bulk allows the apparent Planck scale $M_P$ to
be much larger than the true dynamical scale of gravity $M_* \sim
{\rm ~TeV} $. Consider a spacelike region V of extent $r$ on the
3-brane and thickness $l$ in the orthogonal extra dimensions. The
boundary of V consists of components whose surface areas scale as
$r^3 \, l^{(D-5)}$ and $r^2 \, l^{(D-4)}$. The first surface
component is obtained by setting the extra-dimensional coordinates
at their extreme (boundary) values and allowing the coordinates
$\{ x_{1-3} \}$ to vary throughout the intersection of V with the
3-brane. The second is obtained by setting $\{ x_{1-3} \}$ at
their extreme values (i.e., the boundary on the 3-brane) and
letting the extra-dimensional coordinates to vary over a range of
size $l$.

We will apply the covariant bound to a hypersurface $B$ which
corresponds to the {\it second} part of the boundary of $V$, the
one whose area scales as $r^2 \, l^{(D-4)}$. The light sheet $L$
associated with $B$ intersects all of the ordinary matter in $V$
because, by assumption, it is confined on the brane\footnote{If
the matter were {\it not} confined on the brane, we would have to
include the first surface area component (the $r^3 \, l^{(D-5)}$
component) in $B$ to obtain a bound on the matter entropy.}.
Therefore, the entropy on $L$ is simply that of the ordinary
matter in $V$.

Let $V$ have the same shape as the brane, with thickness $l$ of
order $M_*^{-1}$, so that its surface area is of order $r^2$ in
units of $M_*$. It is possible that the brane is thicker than
$M_*^{-1}$, forcing us to use a larger hypersurface with more
entropy density, however it is hard to imagine that the brane
thickness is parametrically larger than the fundamental length
scale.

A related, model-dependent, subtlety is whether the gravitational
pull of the brane (in the extra dimension, say coordinate $x_5$)
can cause the rays from which our lightsheet $L$ is constructed to
reach a caustic and begin to diverge before reaching the center of
the fiducial volume. In this case much of the matter would never
intersect $L$. However, note that $B$ is taken to have extent of
only $l \sim M_*^{-1}$ in the $x_5$ direction. There is likely no
meaning to lengths less than $M_*^{-1}$ (the fundamental scale of
quantum gravity and also the thickness of the brane), in which
case the brane and lightsheet are maximally thin and therefore
indivisible. This suggests that we need not consider rays on
either side of the brane (along the $x_5$ coordinate), and
therefore no focusing towards the brane.

\bigskip
\noindent  I. {\it The covariant bound is violated during the big
bang.}

Let the temperature in the early universe at temperature T. Impose
that the fiducial region V saturate the holographic bound, so $r$
satisfies \beq \label{saturate} T^3 r^3 \sim M_*^2 r^2~~, \eeq or
\beq r \sim T^{-1} \left( {M_* \over T} \right)^2 ~~.\eeq

Now compare our fiducial volume to a cosmological horizon volume
of size $d_H \sim M_P / T^2$ (assuming radiation domination). The
ratio of $r$ to $d_H$ is \beq \label{horizon} { d_H \over r}
~\sim~ {T \over M_*}{M_P \over M_*} ~\sim~ {T \over {\rm 10^{-4}
\, eV}} ~~~.\eeq For the matter-dominant epoch, the horizon
distance is given as $d_H\sim \left({M_P/T_d^2}\right)\left({T_d/
T}\right)^{3/2}$, where $T_d\simeq 10~{\rm eV}$ is the onset
temperature of matter domination. The ratio then becomes
\begin{equation}
\label{dr} {d_H\over r}\sim {M_P\over M_*}\left({T^3\over M_*^2
T_d}\right)^{1/2} \sim \left({T\over 10^{-2}~{\rm
eV}}\right)^{3/2}.
\end{equation}
We find  that for any temperature higher than ${\rm 10^{-2} \,
eV}$ the causal horizon contains more degrees of freedom than are
allowed according to the CB applied to the fundamental theory.

Our understanding of thermodynamics and statistical physics is
based on counting states. If the CB is correct, the early universe
in the brane worlds under consideration will likely not obey the
usual laws of thermal physics at temperatures $>{\rm 10^{-2} \,
eV}$. This makes our understanding of nucleosynthesis and the
microwave background problematic.

In order that our thermodynamic description of nucleosynthesis (at
$T \sim 10$ MeV) not be invalidated by holography, we find that
$M_* > 10^4$ TeV. (This bound is reduced slightly from
(\ref{horizon}) when prefactors in the expressions for the entropy
density and horizon size are included.)

The maximum violation of the covariant entropy bound can be
deduced by considering a thermal region of temperature $T$ and
extent $r \sim M_P / T^2$ on the 3-brane (the largest it could be
and still fit in a cosmological horizon). Comparing its entropy $S
\sim r^3 T^3 \sim (M_P / T)^3$ to its surface area $r^2$, and
taking the maximum temperature $T \sim M_*$, we see that the
entropy bound can be violated by a factor of $(M_P / M_*) \sim
10^{16}$ for TeV gravity.

\bigskip
\noindent  II. {\it The covariant bound is violated by supernova
cores.}

Consider the supernova of a star of mass $M >  8 M_\odot$, which
is powered by the collapse of an iron core and leads to neutron
star or black hole formation. In this process the entropy of the
collapsed neutron star is of order one per nucleon, so the total
entropy is roughly $10^{57}$. The radius of the core is a few to
ten kilometers, so that its area ($10^{12}~{\rm cm}^2$) in $M_*$
units is only $10^{46}$, where again we take a fiducial volume of
thickness just greater than that of the brane. (As in the
cosmological case the degrees of freedom we are counting are all
confined to the brane.) Unless $M_* > 10^6$ TeV there is a
conflict between the usual thermodynamic description of supernova
collapse and the holographic entropy bound.

\bigskip
\noindent III. {\it Black hole entropy bound vs. covariant bound}

Susskind \cite{Susskind:1994vu} imagines a process in which a
thermodynamic system is converted into a black hole by collapsing
a spherical shell around it. Using the GSL, one obtains a bound on
the entropy of the system: $S_{matter} \leq A/4$, where $A$ is the
area of the black hole formed. This is a {\it weaker} conjecture
than the covariant bound, and has considerable theoretical support
\cite{Susskind:1994vu,Bousso:2002ju,Bekenstein:ur,Bekenstein:ax}.
In the application of the CB we are free to choose the
hypersurface B, as long as its lightsheet intersects all of the
matter whose entropy we wish to bound, whereas in the black hole
bound the area which appears is that of the black hole which is
formed. The black hole entropy bound is sensitive to the dynamics
of horizon formation.

In TeV gravity scenarios, the black hole size on the 3-brane is
controlled by the apparent Planck scale $M_P = 10^{19}$ GeV. The
extent of the horizon in the perpendicular directions off the
brane depends on the model, unless the hole is very small.

In ADD worlds, the horizon of an astrophysical black hole likely
extends to the boundary of the compact extra dimensions. As
discussed in \cite{Argyres:1998qn}, large black holes have
geometry $S^2 \times T^{D-4}$, and the horizon includes all of the
extra volume $V_w$. Due to this additional extra-dimensional
volume, the resulting entropy density is the same as in 3+1
dimensions and there is no obvious violation of any bounds.

In RS scenarios, however, black holes are confined to the brane
and have a pancake-like geometry
\cite{Giddings:2000mu,Casadio:2002uv}. The black hole size in the
direction transverse to the brane grows only logarithmically with
the mass $M$. Thus far, no one has computed the Hawking
temperature or entropy of a pancake black hole. In fact, exact
solutions describing these objects have yet to be obtained. Let us
assume, motivated by holography, that the entropy of a pancake
black hole continues to be of order its surface area in units of
$M_*$. The surface area of a large hole is dominated by the $r^3
l^{D-5}$ component, so the black hole entropy bound arising from
the Susskind construction in RS worlds is of the form \beq
\label{Susskind} S < (r M_*)^3~~.  \eeq That is, the upper bound
on the entropy grows with the apparent 3-volume of the region. In
this case the black hole bound is clearly weaker than the
covariant bound, because the surface B used in the application of
the latter is much smaller than the area of the pancake hole.
Interestingly, (\ref{Susskind}) is the same result one would have
obtained naively from $D = 4$ quantum field theory in the absence
of gravity, with ultraviolet cutoff $M_*$!

Assuming that the entropy of a pancake black hole is given by its
area in $M_*$ units, the relationship between entropy and temperature
implies that the Hawking temperature is
greatly enhanced: \beq \label{Hawking} T ~=~ {dM \over dS} ~\sim~
{1 \over r} \left({M_P \over M}\right) \left( {M_P \over M_*}
\right)^3. \eeq For example, a solar mass black hole would have
temperature $T \sim$ eV if $M_* \sim$ TeV. However,
(\ref{Hawking}) should be treated with caution. The calculation of
the Hawking temperature for a pancake black hole is rather
confusing - an observer on the brane would seem to obtain the
usual result, while an observer in the normal direction might find
a different temperature in radiated gravitons, depending on the
precise (currently unknown) nature of the horizon in the normal
coordinate.

\bigskip

\noindent {\bf Discussion}

Our results can be interpreted in two ways, depending on how one
views holography and related entropy bounds.

It seems likely that holography is a deep result of quantum
gravity, relating geometry and information in a new way
\cite{Bousso:2002ju}. If so, it provides important constraints on
extra dimensional models. Our analysis shows that the ordinary
thermodynamic treatment of nucleosynthesis and supernovae are in
conflict with the covariant bound. In other words, brane worlds
obeying holography do not reproduce the observed big bang thermal
evolution or stellar collapse. Exactly what replaces the usual
behavior is unclear - presumably it is highly non-local - but the
number of degrees of freedom is drastically less than in the
thermodynamic description.

An alternative point of view is to regard brane worlds as a
challenge to holography. If such worlds exist they have the
potential to violate the entropic bounds by arbitrarily large
factors. However, it must be noted that the basic dynamical
assumptions underlying the scenarios (that the 3-brane and bulk
geometry arise as a ground state of quantum gravity) have never
been justified. All violations discussed here require a hierarchy
between $M_P$ and $M_*$, or equivalently that the
extra-dimensional volume factor $V_w = \int d^{D-4}x ~\sqrt{-g_{(D-4)}}  $  exceed its
``natural'' size $\sim M_*^{-(D-4)}$.

Finally, we note that the brane, or whatever confines matter to 3
spatial dimensions, is absolutely necessary for these entropy
violations. Without the brane, matter initially in a region with
small extent in the extra $(D-4)$ dimensions will inevitably
spread out due to the uncertainty principle. For ordinary matter
in classical general relativity, in the absence of branes, Wald
and collaborators \cite{Flanagan:1999jp} have proven the covariant
entropy bound subject to some technical assumptions.

\section*{Acknowledgements}
\noindent

The work of D.K.H. is supported by KRF PBRG 2002-070-C00022.
The work of S.H. was supported in part
under DOE contract DE-FG06-85ER40224 and by the NSF through
through the USA-Korea Cooperative Science Program, 9982164.

\bigskip


\end{document}